\documentclass[aps,prd,preprint,nofootinbib]{revtex4-1}
\pdfoutput=1
\usepackage{graphicx}
\usepackage{psfrag}
\usepackage{float}
\usepackage{subfigure}
\usepackage{array}
\usepackage{graphicx}
\usepackage{amsmath}
\usepackage{amssymb}
\usepackage{mathrsfs}
\usepackage{bm}
\usepackage{slashed}
\usepackage{threeparttable}
\usepackage{multirow}
\usepackage[colorlinks, linkcolor=black, anchorcolor=black, citecolor=black]{hyperref}
\usepackage[amsmath,thmmarks,hyperref]{ntheorem}
\usepackage{soul}
\allowdisplaybreaks[4]

\begin{document}

\title{FCNC processes of charmed hadrons with invisible scalar}
\author{Chao-Qiang Geng, Geng Li\footnote[1]{ligeng@ucas.ac.cn}\\}
\affiliation{School of Fundamental Physics and Mathematical Sciences, Hangzhou Institute for Advanced Study, UCAS, 310024 Hangzhou, China\\
	University of Chinese Academy of Sciences, 100190 Beijing, China
	\vspace{0.6cm}} 

\baselineskip=20pt

\begin{abstract}
We study the invisible scalar of $S$ in charmed meson and baryon decays with flavor-changing neutral currents (FCNCs) based on the model-independent effective Lagrangian between the quarks and invisible scalar. From the bounds of the coupling constants extracted from the recent BES III experiment on the decay of $D^{0}\to\pi^{0}\bar\nu\nu$, we predict that the decay branching ratios of $D^{+}\to\pi^{+}S$, $D_s^+\to K^+S$, $\Lambda_c\to pS$, $\Xi_c^{0(+)}\to \Sigma^{0(+)}S$, and $\Xi_c^0\to\Lambda S$ can be as large as $10.6,~2.53,~2.39,~0.963,~5.77$, and $2.92\times10^{-4}$ with $m_{_S}\approx 1.1~{\rm GeV}$, respectively. Some of these decay modes are accessible to the ongoing experiments, such as BES III, LHCb and Belle II, as well as the future ones, such as FCC-ee. 
\end{abstract}

\maketitle

\section{Introduction}
The flavor-changing neutral current (FCNC) processes of hadrons have attracted more and more attentions, as they provide windows to search for new physics (NP) beyond the standard model (SM). 

At quark level, the decays associated with $s\to d$ and $b\to s(d)$ transitions plus the neutrino ($\nu$) and anti-neutrino ($\bar \nu$) in the final states have been extensively studied. For example, the KOTO experiment at J-PARC~\cite{KOTO:2018dsc} has obtained the upper bound on the decay branching ratio ($\mathcal B$) of $K_L\to\pi^0\bar\nu\nu$ to be $\mathcal {B}(K_L\to\pi^0\bar\nu\nu)_{\rm KOTO}<3.0\times 10^{-9}$ at $90\%$ confidence level (C.L.). The NA62 experiment at CERN~\cite{NA62:2022zos} has measured the decay of $K^+\to \pi^+\bar\nu\nu$, given by $\mathcal {B}(K^+\to\pi^+\bar\nu\nu)_{\rm NA62}=(11.0^{+4.0}_{-3.5}({\rm stat})\pm 0.3(\rm syst))\times 10^{-11}$ at $68\%$ C.L. In addition, the E949 experiment at BNL~\cite{E949:2008btt} has found that $\mathcal {B}(K^+\to\pi^+\bar\nu\nu)_{\rm E949}= (17.3^{+11.5}_{-10.5})\times 10^{-11}$. These results are close to the SM predictions of $\mathcal {B}(K_L\to\pi^0\bar\nu\nu)_{\rm SM} = (3.4 \pm 0.6)\times10^{-11}$ and $\mathcal {B}(K^+\to\pi^+\bar\nu\nu)_{\rm SM}= (8.4\pm1.0)\times 10^{-11}$~\cite{Buras:2015qea}, respectively, indicating that a very tiny space is left for NP in the $s\to d$ transition. Whereas for the $b\to s(d)$ transitions, so far experiments can only obtain the upper limits on the decay channels associate with $\bar\nu\nu$, which leave some rooms for NP. For example, the upper bounds of $\mathcal B(B\to K^{(*)}(\pi, \rho)\bar\nu\nu)$ have been given by the CLEO~\cite{CLEO:2000yzg}, BarBar~\cite{BaBar:2004xlo, BaBar:2008wiw, BaBar:2010oqg, BaBar:2013npw}, Belle~\cite{Belle:2007vmd, Belle:2013tnz, Belle:2017oht} and Belle II~\cite{Belle-II:2021rof} collaborations. The strictest limits among them are still larger than the SM predictions~\cite{Bause:2021cna, Du:2015tda}. Theoretically, the effects of the invisible particles with various spins in the decays associated with the $s\to d$ and $b\to s(d)$ transitions have been explored in Refs.~\cite{Berezhiani:1989fs, Bird:2004ts, Bird:2006jd, Kamenik:2011vy, Gninenko:2015mea, Bertuzzo:2017lwt, Barducci:2018rlx, Li:2018hgu, Li:2020dpc, Li:2021sqe, Li:2022tbh}. 

The decays with the $c\to u$ transition are also promising, but there are few experimental researches at present. The first measurement on charmed-hadron decays with $\bar\nu\nu$ in the final states is from the BES III experiment~\cite{BESIII:2021slf}, which is given as $\mathcal B (D^0\to\pi^0\bar\nu\nu)<2.1\times10^{-4}$. As the experiments are ongoing, more relevant decay channels involving both mesons and baryons are expected to be measured in the near future, such as BES III, LHCb~\cite{Cerri:2018ypt}, Belle II~\cite{Belle-II:2018jsg} and the future FCC-ee~\cite{FCC:2018byv, FCC:2018evy, Blondel:2014bra}. In the SM, the decays with the $c\to u$ transition are strongly suppressed by the Glashow-Iliopoulos-Maiani (GIM) mechanism. Once signals of such processes are found in experiments, it will be a clear evidence for NP. 

In the theoretical studies, Refs.~\cite{Bause:2020xzj, Faisel:2020php} provided the upper limits on the branching ratios of $c\to u\bar\nu\nu$ channels with model-independent ways, as well as $Z'$ and leptoquark models. And Refs.~\cite{Burdman:2001tf, Bause:2019vpr, Golz:2021imq, Golz:2022alh} analyzed rare charm hadron decays with the $c\to u l^+l^-$ transition within the SM and beyond. In addition, Ref.~\cite{Su:2020yze} searched for the massless dark photons in the decays with the $c\to u$ transition. Scalar dark matter candidates can be achieved in minimal extensions of the SM~\cite{OConnell:2006rsp, Patt:2006fw}, in which the hidden scalar can mix with the Higgs boson~\cite{Krnjaic:2015mbs, Winkler:2018qyg, Filimonova:2019tuy, Kachanovich:2020yhi}. One of the most famous hidden spin-0 candidates is the axion-like particle (ALP). The QCD-axion was firstly introduced in order to explain the strong-CP problem~\cite{Peccei:1977hh, PhysRevLett.40.223, Wilczek:1977pj}. More recently, ALPs are considered as one of the best-motivated candidates for particle extensions of the SM~\cite{Batell:2009jf, Aditya:2012ay}. The values of their masses are arbitrary~\cite{Izaguirre:2016dfi}, and their interactions with SM fields arise through higher-dimensional (or otherwise suppressed) operators. ALPs within models of hidden sectors can either act as the dark matter candidate itself~\cite{Preskill:1982cy, Abbott:1982af, Dine:1982ah}, or as a mediator through the axion portal~\cite{Nomura:2008ru}. In this work, we consider an invisible scalar $S$  emitted in the FCNC processes of charmed hadrons. In particular, we concentrate on the processes of $D^{0(+)}\to\pi^{0(+)}S$, $D_s^+\to K^+S$, $\Lambda_c\to pS$, $\Xi_c^{0(+)}\to \Sigma^{0(+)}S$, and $\Xi_c^0\to\Lambda S$. 

The paper is organized as follows: In Sec. II, we obtain the SM expectations of the dineutrino modes. In Sec. III, we first construct the model-independent effective Lagrangian, which describes the couplings between the quarks and invisible scalar. We then present the numerical results of the upper limits for the decay branching ratios. The hadronic transition matrix elements are evaluated based on the lattice QCD (LQCD) and modified bag model (MBM). Finally, we give the conclusion in Sec. IV.

\section{The SM predictions of the dineutrino modes}
\begin{figure}[htbp]
	\centering
	\subfigure[~Mesonic FCNC processes]{
		\label{Feyn-16}
		\includegraphics[width=0.45\textwidth]{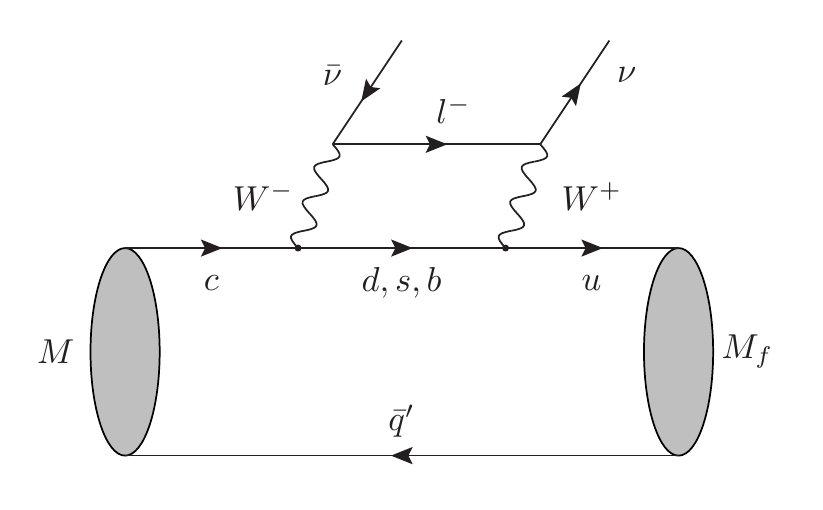}} 
	\hspace{2em}
	\subfigure[~Baryonic FCNC processes]{
		\label{Feyn-13}
		\includegraphics[width=0.45\textwidth]{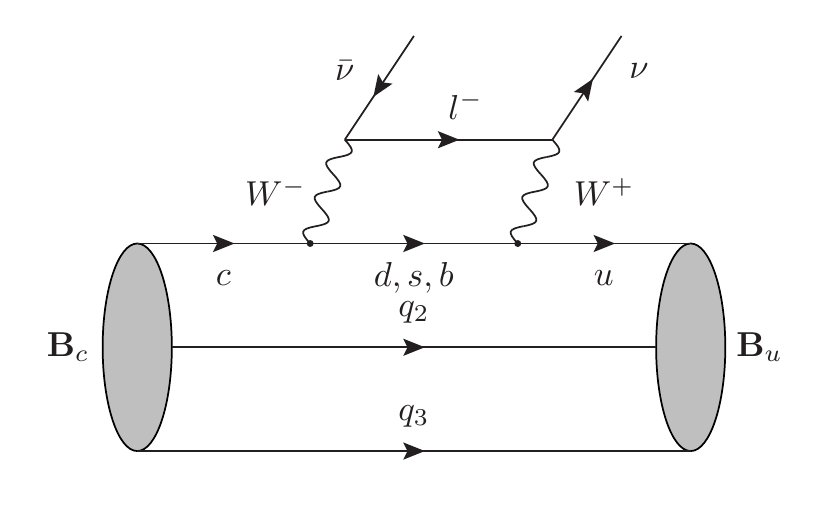}}
	\caption{Feynman diagrams of charmed hadron decays with $\bar\nu\nu$.}
	\label{Feyn}
\end{figure}

In the SM, the FCNC processes of charmed hadrons associated with $\bar\nu\nu$ in the final states are contributed by both short-distance and long-distance effects. As there is no tree-level contribution to the FCNC decays, the first-order short-distance contribution comes from the penguin and box diagrams, as shown in Fig.~\ref{Feyn}, where $M$ and $M_f$ (${\bf B}_c$ and ${\bf B}_u$) represent the initial and final mesons (baryons), $q=c$ and $q_{f}=u$ are initial and final quarks, and $\bar q'$ and $~q_{2(3)}$ are the spectator quarks,  respectively. 
The one-loop Feynman diagrams can be described by the effective Lagrangian, given by~\cite{Inami:1980fz}
\begin{equation}
	\begin{aligned}
		\mathcal L_{c\bar\nu\nu} = \frac{4G_F}{\sqrt{2}}\frac{\alpha}{2\pi\sin^2\theta_W}\sum_{l=e, \mu, \tau}\sum_{q=s,b}V_{cq}V_{uq}X^l(x_q)(\bar u_{_L}\gamma^\mu c_{_{L}})(\bar\nu_{_{lL}}\gamma_\mu\nu_{_{lL}}),
		\label{c-uvv}
	\end{aligned}
\end{equation}
where $G_F$ represents the Fermi coupling constant, $\alpha$ corresponds to the fine structure constant, $\theta_W$ stands for the Weinberg angle, and $V_{ij}$ are the Cabibbo-Kobayashi-Maskawa (CKM) matrix elements.
In Eq.~\eqref{c-uvv}, the masses of quarks are set as $m_d=0,~m_s=0.1~{\rm GeV},~m_b=4.78~{\rm GeV}$~\cite{EuropeanTwistedMass:2014osg}, and $X^l(x_q)$ is related to the Inami-Lim function of $\bar D(x_q, y_l)$, given by~\cite{Inami:1980fz}
\begin{equation}
	\begin{aligned}
		X^l(x_q)&=\frac{\bar D(x_q, y_l)}{2},\\
		\bar D(x_q, y_l) &= \frac{1}{8}\frac{x_qy_l}{x_q-y_l}\left(\frac{y_l-4}{y_l-1}\right)^2\ln y_l + \frac{1}{8}\left[\frac{x_q}{y_l-x_q}\left(\frac{x_q-4}{x_q-1}\right)^2+1+\frac{3}{(x_q-1)^2}\right]\\
		&~~~\times x_q\ln x_q +\frac{x_q}{4}-\frac{3}{8}\left(1+\frac{3}{y_l-1}\right)\frac{x_q}{x_q-1},
		\label{Inami-Lim}
	\end{aligned}
\end{equation}
where $x_q=m_q^2/M_W^2$ and $y_l=m_l^2/M_W^2$ with $m_q$, $m_l$ and $M_W$ being the masses of the quark, lepton and $W$-boson, respectively. Consequently, the transition amplitudes of charmed meson (baryon) decays are given by
\begin{equation}
	\begin{aligned}
		\langle M_f({\bf B}_u)\nu_{_l}\bar\nu_{_l}|\mathcal L_{c\bar\nu\nu}|M({\bf B}_c)\rangle &=\frac{\sqrt{2}G_F\alpha}{4\pi\sin^2\theta_W}\sum_qV^\ast_{cq}V_{uq}X^l(x_q)\langle M_f({\bf B}_u)|\bar c\gamma^\mu(1-\gamma^5)u|M({\bf B}_c)\rangle\\
		&~~~\times \bar u_{\nu_l}\gamma_\mu(1-\gamma^5)v_{\nu_l}.
		\label{amp-vv}
	\end{aligned}
\end{equation}

\begin{figure}[h]
	\centering
	\subfigure[~$D^+\to \pi^+$]{
		\label{FF-D1}
		\includegraphics[width=0.45\textwidth]{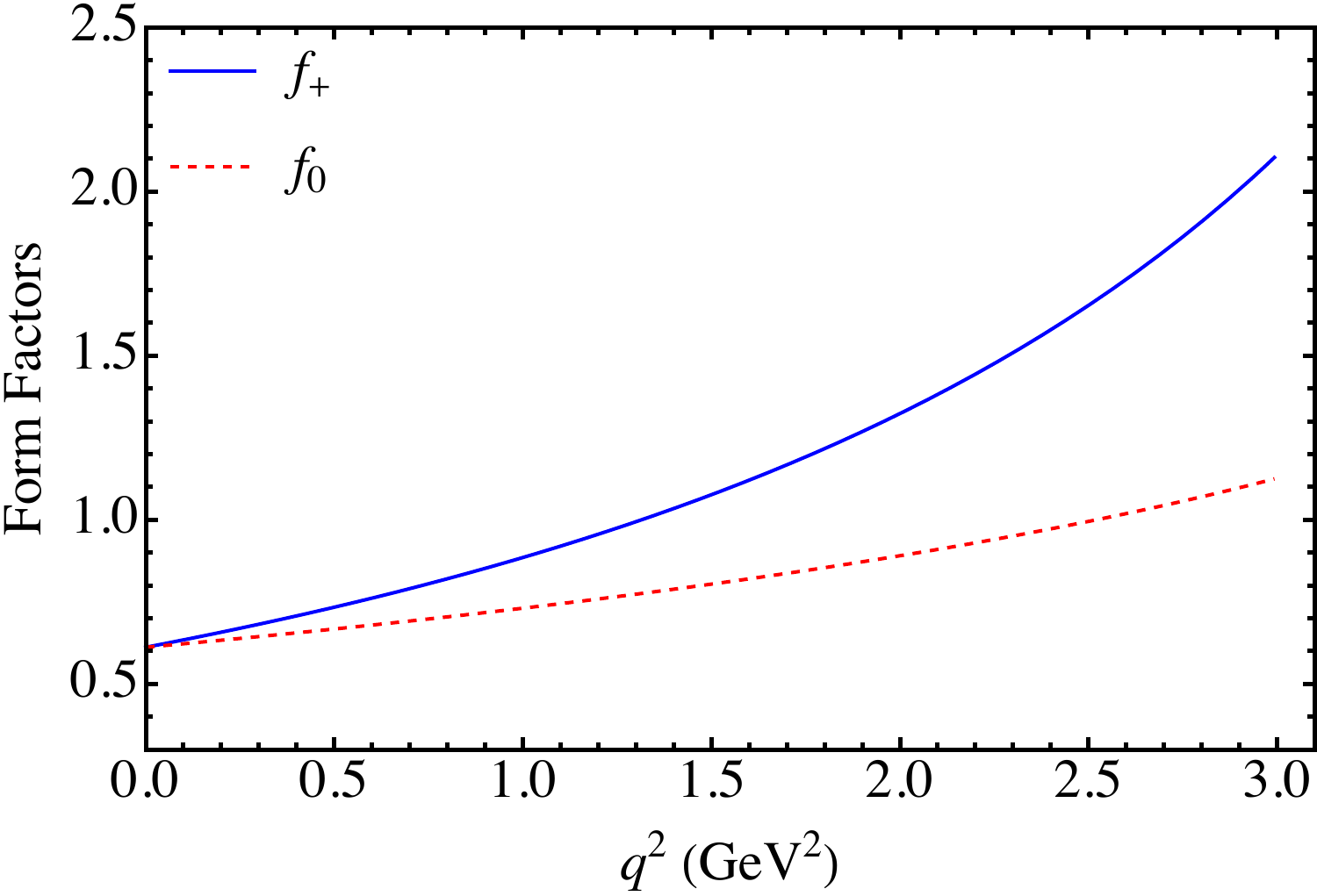}} 
	\hspace{2em}
	\subfigure[~$D^0\to \pi^0$]{
		\label{FF-D0}
		\includegraphics[width=0.45\textwidth]{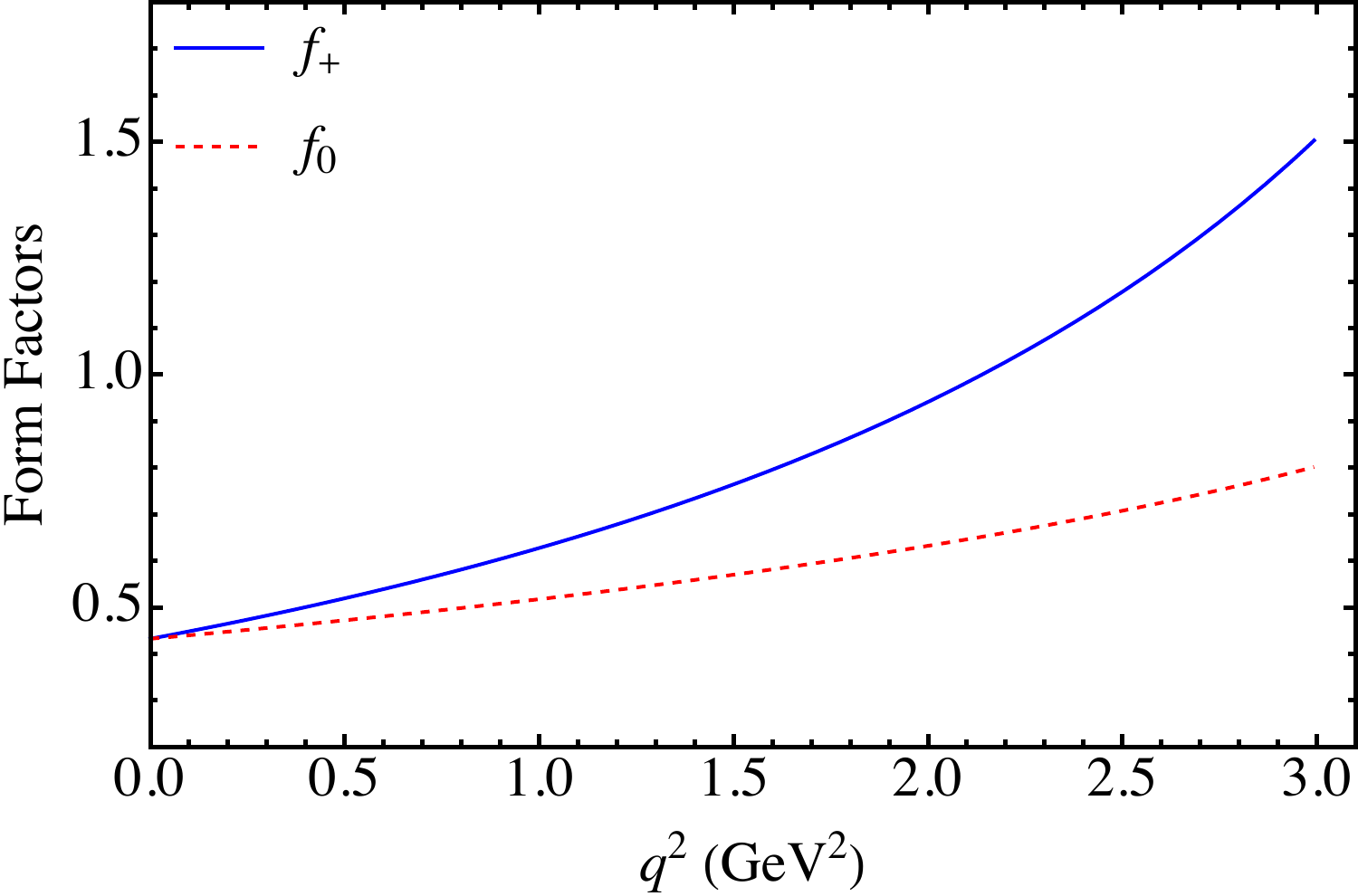}} \\
	\subfigure[~$D_s^+\to K^+$]{
		\label{FF-Ds}
		\includegraphics[width=0.45\textwidth]{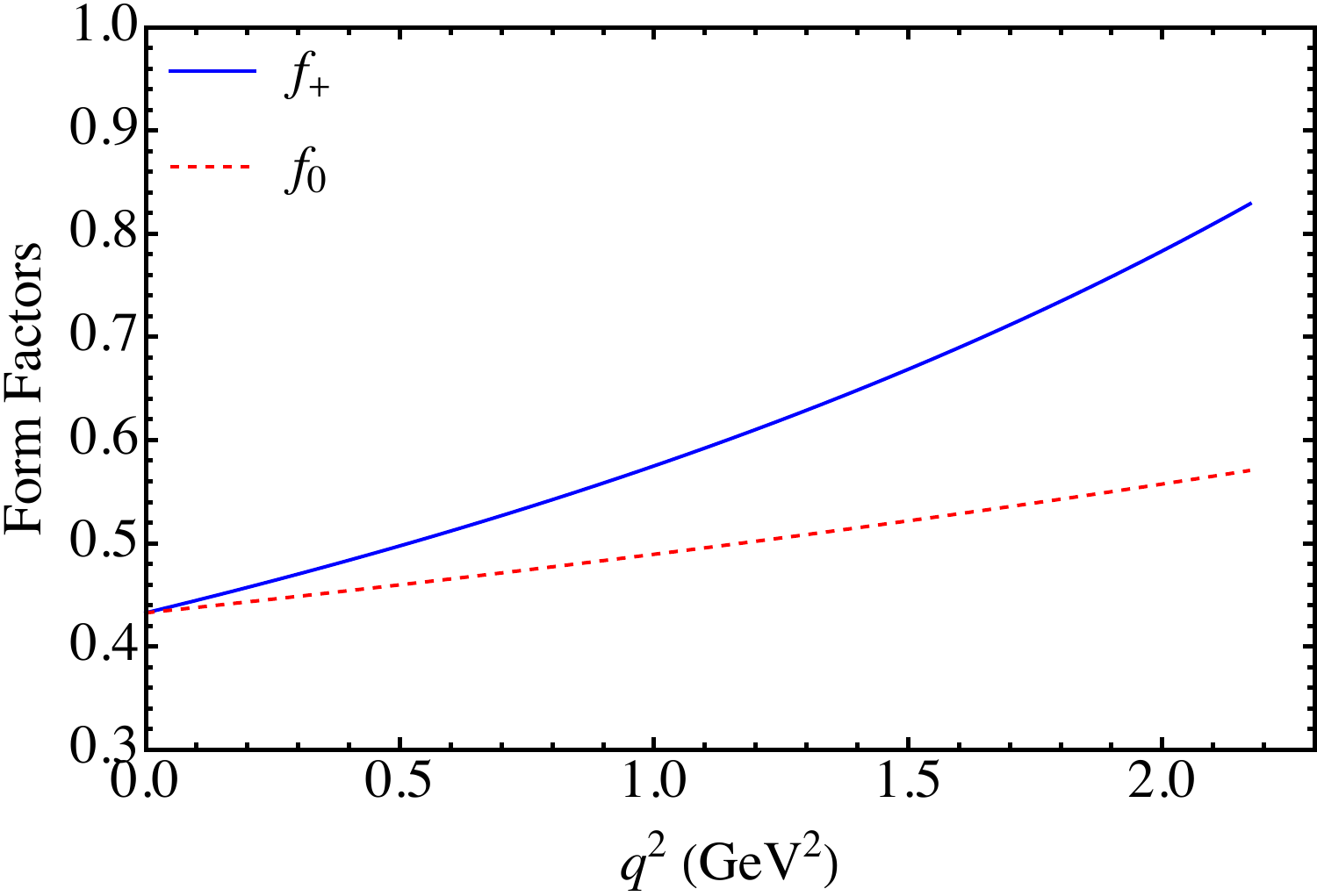}} 
	\caption{Mesonic form factors of (a) $D^+\to \pi^+$, (b) $D^0\to \pi^0$, and (c) $D_s^+\to K^+$.}
	\label{FF-meson}
\end{figure}

The mesonic transition matrix elements in Eq.~\eqref{amp-vv} can be expressed by the form factors (FFs) of $f_0$ and $f_+$, given by
\begin{equation} 
	\begin{aligned}
		\langle M_f^-(P_f)|(\bar q_{_f} q) |M^-(P)\rangle 
		&= \frac{M^2-M_{f}^2}{m_q-m_{q_{_f}}}f_0 (q^2),\\
		\langle M_f^-(P_f)|(\bar q_{_f}\gamma_\mu q) |M^-(P)\rangle 
		&= (P+P_f)_{\mu}f_+ (q^2)+(P-P_f)_{\mu}\frac{M^2-M_{f}^2}{q^2} \big[f_0 (q^2)-f_+ (q^2)\big], 
	\end{aligned}
\end{equation}
where $P$ ($P_f$) and $M$ ($M_f$) are the momentum and mass of the initial (final) hadron, respectively, and $q^2$ corresponds to the momentum transfer. We adopt the numerical results of the $D\to K$ form factors from the LQCD~\cite{Lubicz:2017syv,EuropeanTwistedMass:2014osg}, where the form factors of $D^0 \to\pi^0$ receive additional isospin factors of $f_i \to f_i/\sqrt 2~(i=0,+)$. According to the lattice study of  Ref.~\cite{Koponen:2012di}, we use the same numerical inputs for $D \to\pi$ and $D_s \to K$ besides obvious kinematic replacements. The behaviors of the FFs along with their $q^2$ dependencies are shown in Fig.~\ref{FF-meson}.

\begin{figure}[h]
	\centering
	\subfigure[~$\Lambda_c\to p$]{
		\label{ff-Lcp}
		\includegraphics[width=0.45\textwidth]{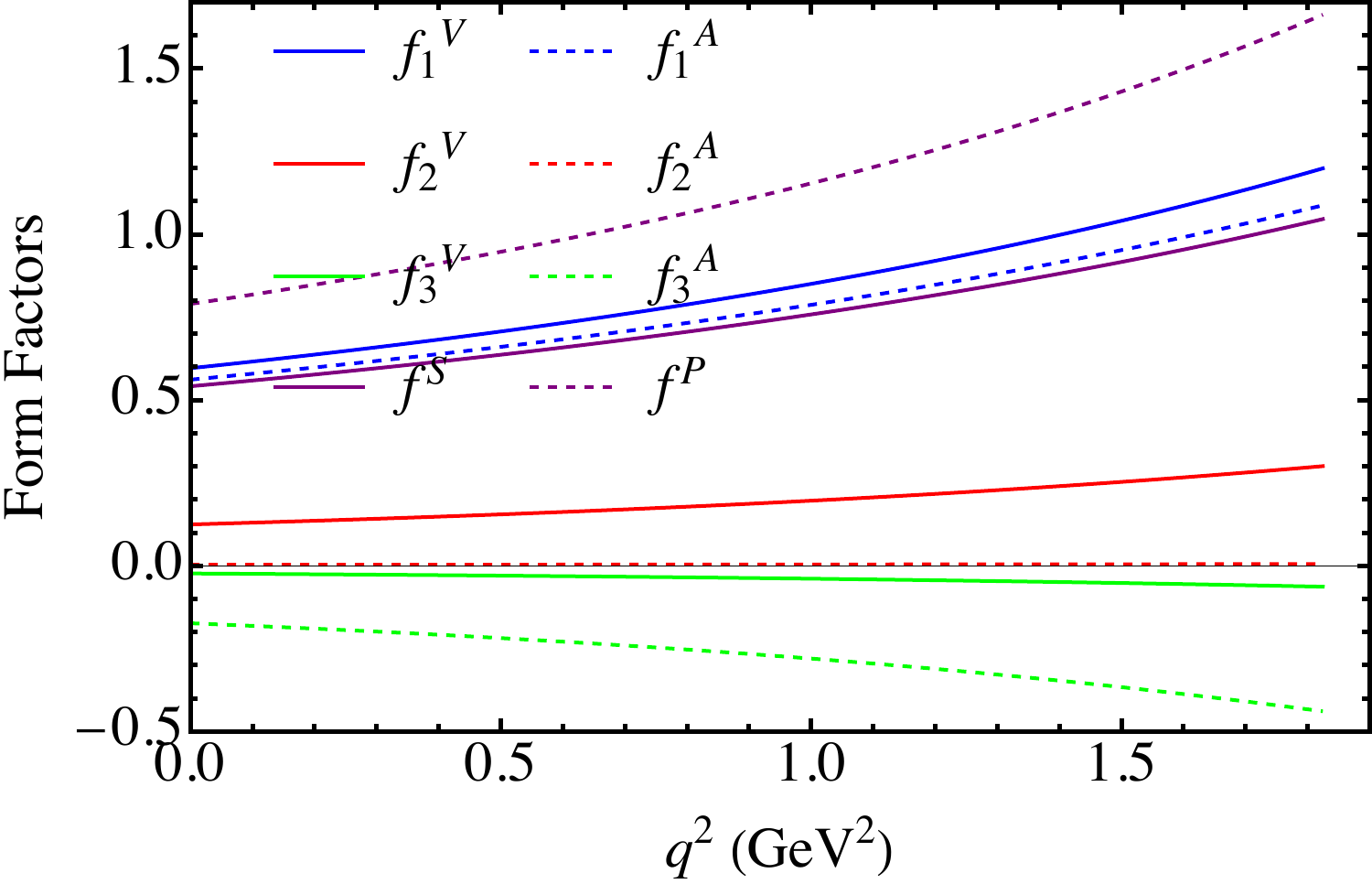}} 
	\hspace{2em}
	\subfigure[~$\Xi_c^{+}\to \Sigma^{+}$]{
		\label{ff-Xc1S1}
		\includegraphics[width=0.45\textwidth]{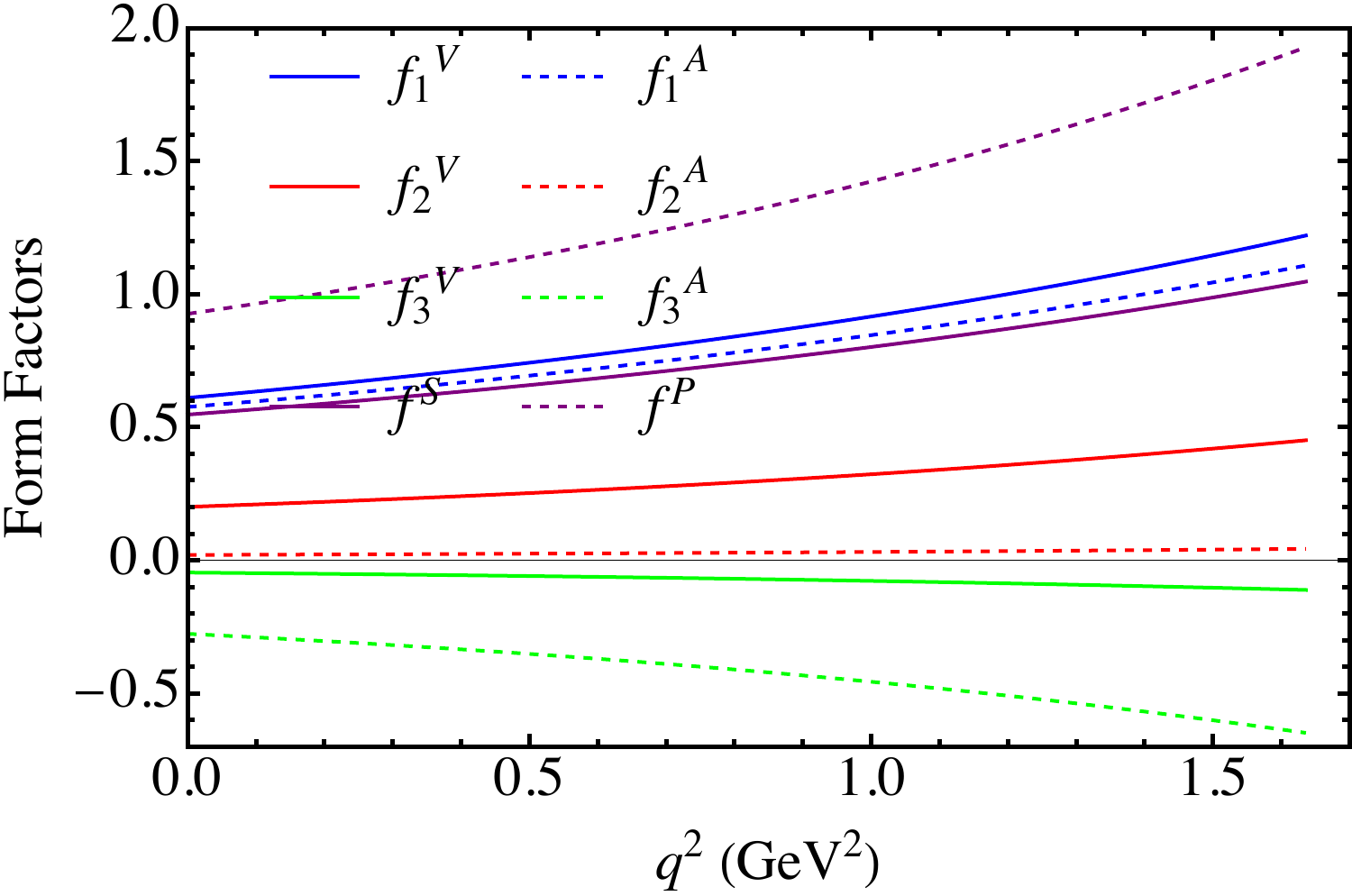}} \\
	\subfigure[~$\Xi_c^{0}\to \Sigma^{0}$]{
		\label{ff-Xc0S0}
		\includegraphics[width=0.45\textwidth]{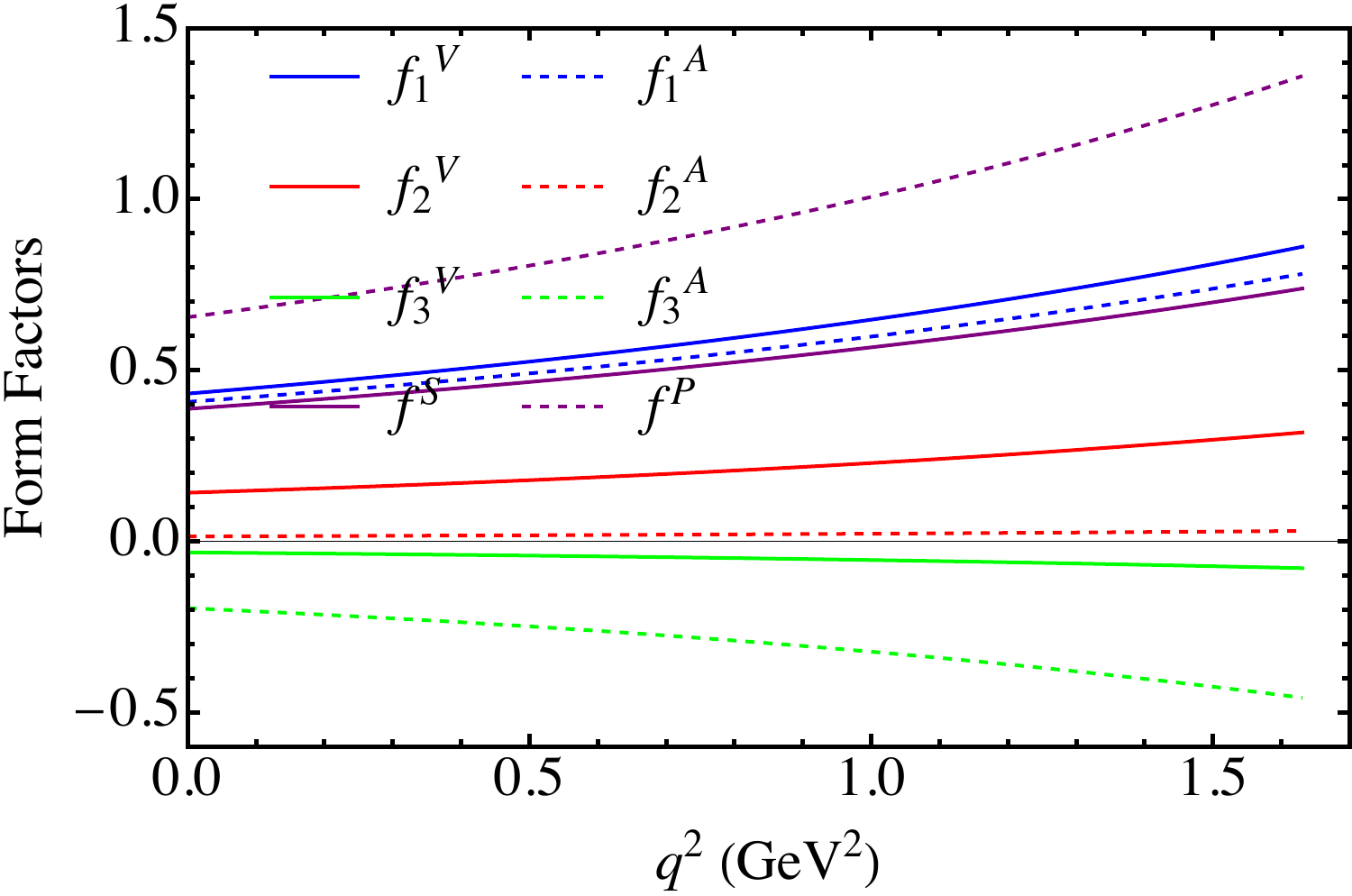}} 
	\hspace{2em}
	\subfigure[~$\Xi_c^0\to \Lambda$]{
		\label{ff-Xc0L}
		\includegraphics[width=0.45\textwidth]{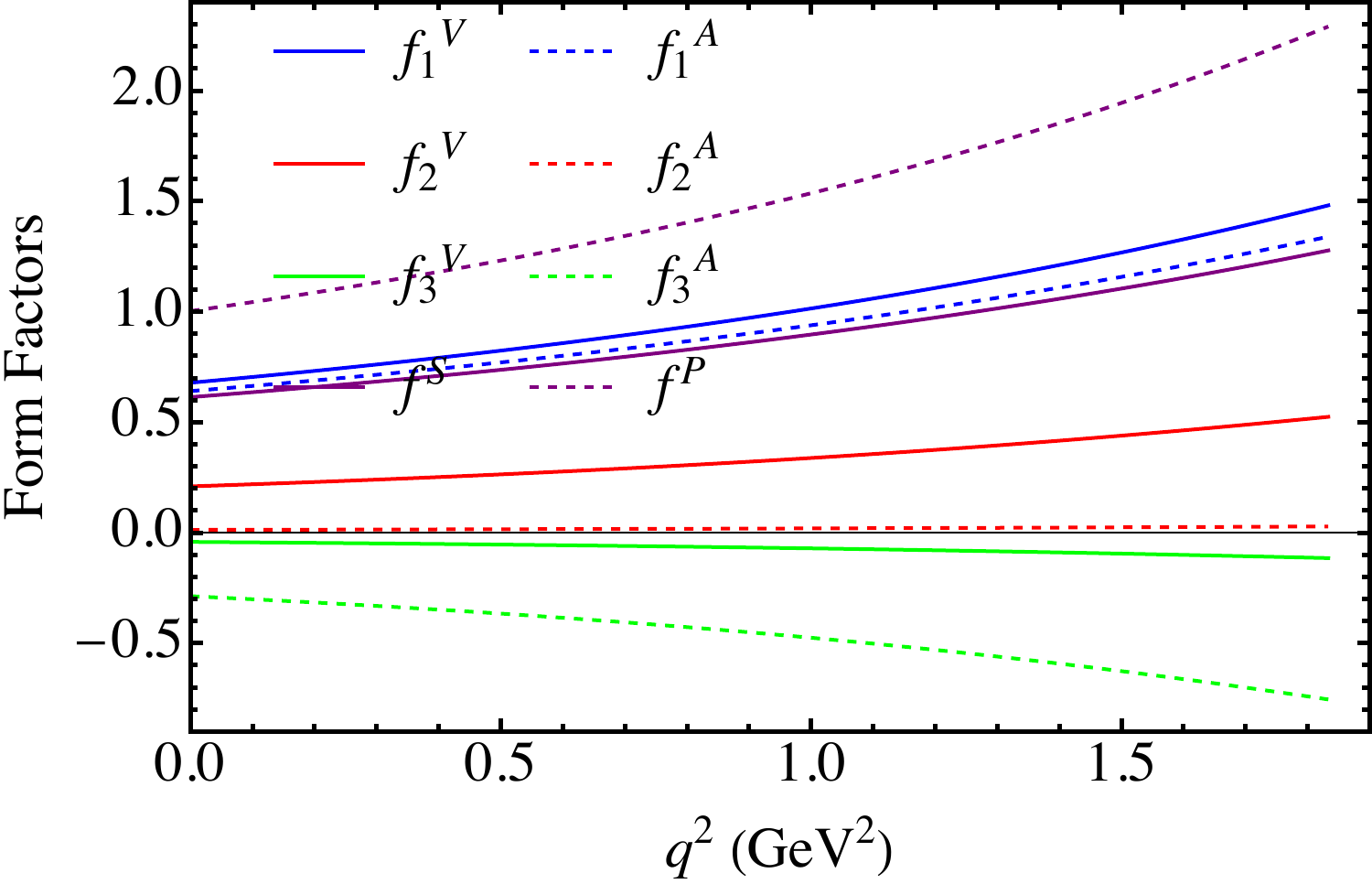}} 
	\caption{Baryonic form factors of (a) $\Lambda_c\to p$, (b) $\Xi_c^{+}\to \Sigma^{+}$, (c) $\Xi_c^{0}\to \Sigma^{0}$, and (d) $\Xi_c^0\to \Lambda$.}
	\label{FF-baryon}
\end{figure}

The baryonic transition matrix elements in Eq.~\eqref{amp-vv} can be parameterized by the FFs of $f_i^{V,A}~(i=1,2,3)$, $f^S$ and $f^P$, defined by
\begin{equation}
	\begin{aligned}
		\langle {\bf B}_u(P_f)|(\bar q_{f}\gamma_\mu q) |{\bf B}_c(P)\rangle 
		&=\bar u_{ _{{\bf B}_u}} (P_f)\left[\gamma_\mu f^V_1(q^2) + i\sigma_{\mu\nu} \frac{q^\nu}{M} f^V_2(q^2)+\frac{q_\mu}{M}f^V_3(q^2)\right]u_{_{{\bf B}_c}}(P),\\
		\langle {\bf B}_u(P_f)|(\bar q_{f} q) |{\bf B}_c(P)\rangle 
		&=\bar u_{_{{\bf B}_u}} (P_f)f^S(q^2)u_{_{{\bf B}_c}}(P),\\
		\langle {\bf B}_u(P_f)|(\bar q_{f}\gamma_\mu \gamma^5 q) |{\bf B}_c(P)\rangle
		&=\bar u_{_{{\bf B}_u}} (P_f)\left[\gamma_\mu f^A_1(q^2) + i\sigma_{\mu\nu} \frac{q^\nu}{M} f^A_2(q^2)+\frac{q_\mu}{M}f^A_3(q^2)\right]\gamma^5u_{_{{\bf B}_c}}(P),\\
		\langle {\bf B}_u(P_f)|(\bar q_{f} \gamma^5 q) |{\bf B}_c(P)\rangle 
		&=\bar u_{_{{\bf B}_u}} (P_f)f^P(q^2)\gamma^5 u_{_{{\bf B}_c}}(P). 
		\label{ffb}
	\end{aligned}
\end{equation}
These elements are evaluated by the MBM. The MBM works well for the heavy baryonic decays with more details found in Refs.~\cite{Geng:2020ofy,Geng:2021sxe,Liu:2021rvt,Liu:2022bdq,Liu:2022pdk,Li:2022tbh}. Here, we only give our numerical results of the FFs as functions of $q^2$ directly in Fig.~\ref{FF-baryon}. 

The decay branching ratio is given by
\begin{equation}
	\mathcal {B} = \frac{1}{512  \pi^3 M^3 \Gamma}\int\frac {dq^2}{q^2} \lambda^{1/2}(M^2, q^2, M_f^2)\lambda^{1/2}(q^2, m_1^2, m_2^2)\int d\cos\theta|\overline{\mathcal  M}|^2,
	\label{ps3}
\end{equation}
where $\lambda(x, y, z)= x^2 + y^2 +z^2 -2xy-2xz -2yz$ is the K${\rm \ddot a}$llen function, $m_{1(2)}$ corresponds to the mass of the (anti-)neutrino, $\theta$ is the phase space angle, $\Gamma$ represents the total width of the initial hadron, and $|\overline{\mathcal  M}|^2$ is the squared amplitude averaged over the initial polarization and summed over the final polarizations. As a result, we obtain the short-distance contributions of charmed hadronic FCNC processes with $\bar\nu\nu$, which are given in Table~\ref{SM}. 
\begin{table}[h]
	\setlength{\tabcolsep}{2em}
	\caption{The branching ratios of charmed hadronic FCNC processes with $\bar\nu\nu$.}
	\centering
	\begin{tabular*}{\textwidth}{@{}@{\extracolsep{\fill}}ccc}
		\hline\hline
		Modes&This work&Ref.~\cite{Burdman:2001tf}\\
		\hline
		$D^0  \to \pi^0\bar \nu \nu $&$ 5.05\times 10^{-17}$&$5.0\times 10^{-16}$   \\
		$D^+  \to \pi^+\bar \nu \nu $&$ 2.56\times 10^{-16}$&$1.2\times 10^{-15}$ ($6.9\times 10^{-16}$)	\\
		$D_s^+  \to K^+\bar \nu \nu $&$ 4.12\times 10^{-17}$ & \\
		$\Lambda_c\to p\bar\nu\nu$&$1.40\times 10^{-16}$	& \\
		$\Xi_c^+\to \Sigma^+\bar\nu\nu$&$3.09\times 10^{-16}$ & \\
		$\Xi_c^0\to \Sigma^0\bar\nu\nu$&$5.14\times 10^{-17}$ & \\
		$\Xi_c^0\to \Lambda\bar\nu\nu$&$1.69\times 10^{-16}$ & \\ 
		\hline\hline
		\label{SM}
	\end{tabular*}
\end{table}
In the table, we also show the short(long)-distance contributions from Ref.~\cite{Burdman:2001tf} for comparison. Notably, the short-distance effects are sensitive to the mass parameters of $m_{s,b}$ and updated form factors in the hadron transition matrix elements, which lead to an order of magnitude difference in the results of the branching ratios. The decay branching ratios associated with $c\to u\bar\nu\nu$ are $\mathcal O(10^{-17})$ to $\mathcal O(10^{-15})$, which are strongly suppressed. Both the short- and long-distance contributions can be safely ignored compared to the contribution of the new invisible scalar. It is clear that a NP signal can be easily identified due to the tiny SM background.

\section{Processes with invisible scalar}
\begin{figure}[htbp]
	\centering
	\subfigure[~Mesonic FCNC processes]{
		\label{Feyn-17}
		\includegraphics[width=0.45\textwidth]{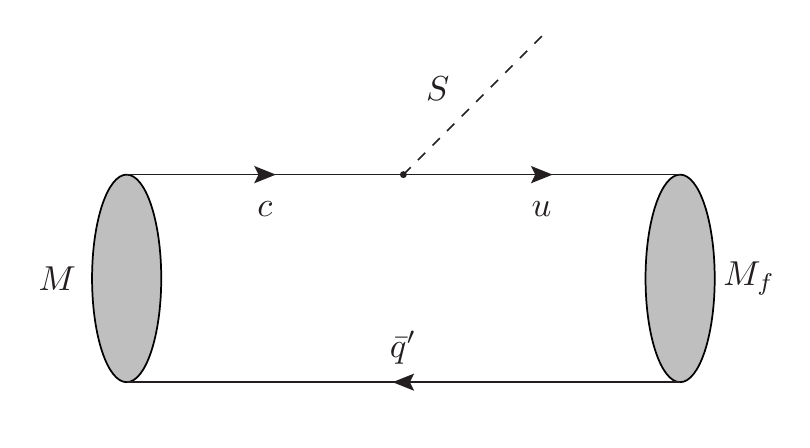}} 
	\hspace{2em}
	\subfigure[~Baryonic FCNC processes]{
		\label{Feyn-14}
		\includegraphics[width=0.45\textwidth]{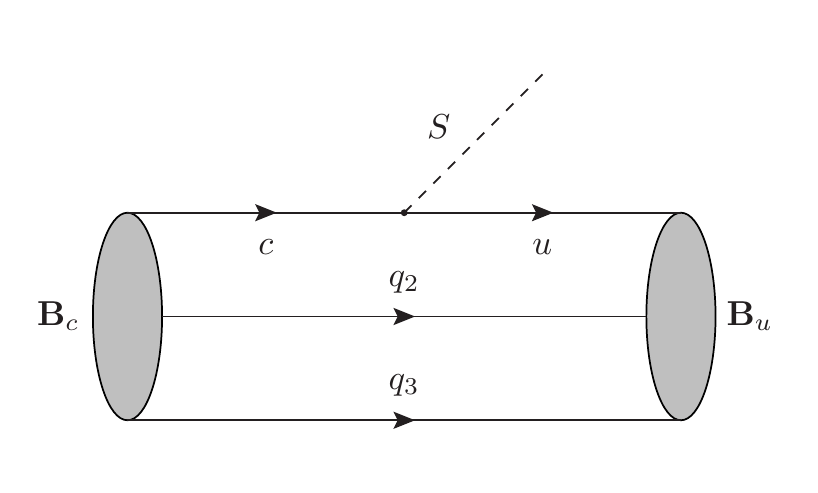}}
	\caption{Feynman diagrams of charmed hadron decays with the invisible scalar of $S$.}
	\label{Feyn-S}
\end{figure}

We now consider the processes discussed in the previous section by replacing the dineutrino with a new invisible scalar, $S$. The Feynman diagrams are shown in Fig.~\ref{Feyn-S}.
The coupling vertex can be described by the effective Lagrangian, given by
\begin{equation}
	\begin{aligned}
		\mathcal L_{scalar}=\sum_{i=1}^{4}{g_{_{Si}} Q_i},
		\label{eq1}
	\end{aligned}
\end{equation}
with
\begin{equation}
	\begin{aligned}
		Q_1&=m_{_S} g_{_{S1}} (\bar q_{_f} q) S~~~~~~~~~~Q_2=m_{_S} g_{_{S2}} (\bar q_{_f} \gamma^{5} q) S \\
		Q_3&=g_{_{S3}} (\bar q_{_f}\gamma_{\mu}q)(i\partial^\mu S) ~~~~~Q_4=g_{_{S4}} (\bar q_{_f} \gamma_{\mu}\gamma^{5}q)(i\partial^\mu S), 
		\label{eq1}
	\end{aligned}
\end{equation}
where $g_{_{Si}}$ are the effective coupling constants. The decay amplitudes of the $0^-\to 0^-$ mesonic decays can be written as
\begin{equation}
	\begin{aligned}
		\langle M_fS|\mathcal L_{scalar}|M \rangle & =m_{_S}g_{_{S1}}\langle M_f^-|(\bar q_{_f} q)|M^-\rangle+g_{_{S3}}	\langle M_f^-|(\bar q_{_f}\gamma_\mu q) |M^-\rangle P_{_S}^\mu\\
		&=g_{_{S1}} \mathcal {T}_1+g_{_{S3}} \mathcal {T}_3. 
		\label{eq1.2}
	\end{aligned}
\end{equation}
And the baryonic decay amplitudes are given by
\begin{equation}
	\begin{aligned}
		\langle {\bf B}_u S|\mathcal L_{scalar}| {\bf B}_c \rangle & =m_{_S}g_{_{S1}}\langle {\bf B}_u|(\bar q_{_f} q)|{\bf B}_c\rangle+m_{_S}g_{_{S2}}\langle {\bf B}_u|(\bar q_{_f} \gamma^5 q)|{\bf B}_c\rangle+g_{_{S3}}	\langle {\bf B}_u|(\bar q_{_f}\gamma_\mu q) |{\bf B}_c\rangle P_{_S}^\mu\\
		& ~~~+g_{_{S4}}	\langle {\bf B}_u|(\bar q_{_f}\gamma_\mu \gamma^5 q) |{\bf B}_c\rangle P_{_S}^\mu \\
		&=g_{_{S1}} \mathcal {T}_1+g_{_{S2}} \mathcal {T}_2+g_{_{S3}} \mathcal {T}_3+g_{_{S4}} \mathcal {T}_4,
		\label{eq1.2}
	\end{aligned}
\end{equation}
where $\mathcal T_i$ are amplitudes other than the effective coupling coefficients, $m_{_S}$ ($P_{_S}$) is the mass (momentum) of the invisible scalar. By integrating the two-body phase space, the decay widths of mesonic and baryonic FCNC decays with the invisible scalar are found to be
\begin{equation}
	\begin{aligned}
		\Gamma(M\to M_fS)=&\frac{1}{16\pi M^3 } \lambda^{1/2}(M^2,M_f^2,m_{_S}^2) \bigg\{ m_{_S}^2 g_{_{S1}}^2 \langle M_f^-|(\bar q_{_f} q)|M^-\rangle^*\langle M_f^-|(\bar q_{_f} q)|M^-\rangle \\
		&+ g_{_{S3}}^2 \langle M_f^-|(\bar q_{_f} \gamma_\nu q)|M^-\rangle^*\langle M_f^-|(\bar q_{_f} \gamma_\mu q)|M^-\rangle P_{_S}^\nu P_{_S}^\mu \\
		&+m_{_S} g_{_{S1}} g_{_{S3}}^* \langle M_f^-|(\bar q_{_f} \gamma_\nu q)|M^-\rangle^*\langle M_f^-|(\bar q_{_f}  q)|M^-\rangle P_{_S}^\nu +h.c. \bigg\}, 
		\label{eq3}
	\end{aligned}
\end{equation}
and
\begin{equation}
	\begin{aligned}
		\Gamma({\bf B}_c\to {\bf B}_u S)&=\frac{1}{16\pi M^3 } \lambda^{1/2}(M^2,M_f^2,m_{_S}^2) \bigg\{ m_{_S}^2 g_{_{S1}}^2 \langle {\bf B}_u|(\bar q_{_f} q)|{\bf B}_c\rangle^*\langle {\bf B}_u|(\bar q_{_f} q)|{\bf B}_c\rangle \\
		&~~~+m_{_S}^2 g_{_{S2}}^2 \langle {\bf B}_u|(\bar q_{_f} \gamma^5 q)|{\bf B}_c\rangle^*\langle {\bf B}_u|(\bar q_{_f} \gamma^5 q)|{\bf B}_c\rangle \\
		&~~~+ g_{_{S3}}^2 \langle {\bf B}_u|(\bar q_{_f} \gamma_\nu q)|{\bf B}_c\rangle^*\langle {\bf B}_u|(\bar q_{_f} \gamma_\mu q)|{\bf B}_c\rangle P_{_S}^\nu P_{_S}^\mu \\
		&~~~+ g_{_{S4}}^2 \langle {\bf B}_u|(\bar q_{_f} \gamma_\nu \gamma^5 q)|{\bf B}_c\rangle^*\langle {\bf B}_u|(\bar q_{_f} \gamma_\mu \gamma^5 q)|{\bf B}_c\rangle P_{_S}^\nu P_{_S}^\mu \\
		&~~~+m_{_S} g_{_{S1}} g_{_{S3}}^* \langle {\bf B}_u|(\bar q_{_f} \gamma_\mu q)|{\bf B}_c\rangle^*\langle {\bf B}_u|(\bar q_{_f}  q)|{\bf B}_c\rangle P_{_S}^\mu +h.c. \\
		&~~~+m_{_S} g_{_{S2}} g_{_{S4}}^* \langle {\bf B}_u|(\bar q_{_f} \gamma_\mu \gamma^5 q)|{\bf B}_c\rangle^*\langle {\bf B}_u|(\bar q_{_f} \gamma^5  q)|{\bf B}_c\rangle P_{_S}^\mu +h.c. \bigg  \},
		\label{eq3}
	\end{aligned}
\end{equation}
respectively. We see that in the $0^- \to 0^-$ mesonic processes, only $\Gamma_{11,33}$ and their interference term $\Gamma_{13}$ make contributions. While in the baryonic ones, all operators give contributions, in which non-zero interference terms are $\Gamma_{13}$ and $\Gamma_{24}$. As the Lagrangian is the sum of several operators, the partial width can be written as
\begin{equation}
	\begin{aligned}
		\Gamma =\int {dPS_2 \big(\sum_{j} g_{_{Sj}}\mathcal{T}_j\big)^{\dagger} \big(\sum_{i} g^{ }_{_{Si}} \mathcal{T}_i \big) }=\sum_{ij}g_{_{Sj}}g_{_{Si}}\widetilde\Gamma_{ij}, 
		\label{eq2}
	\end{aligned}
\end{equation}
Here, we have defined $\widetilde\Gamma_{ij}=\int dPS_2 \mathcal{T}_{j}^\dagger \mathcal{T}_{i}$, which are independent of the coefficients. 

\begin{figure}[htb]
	\centering
	\includegraphics[width=0.45\textwidth]{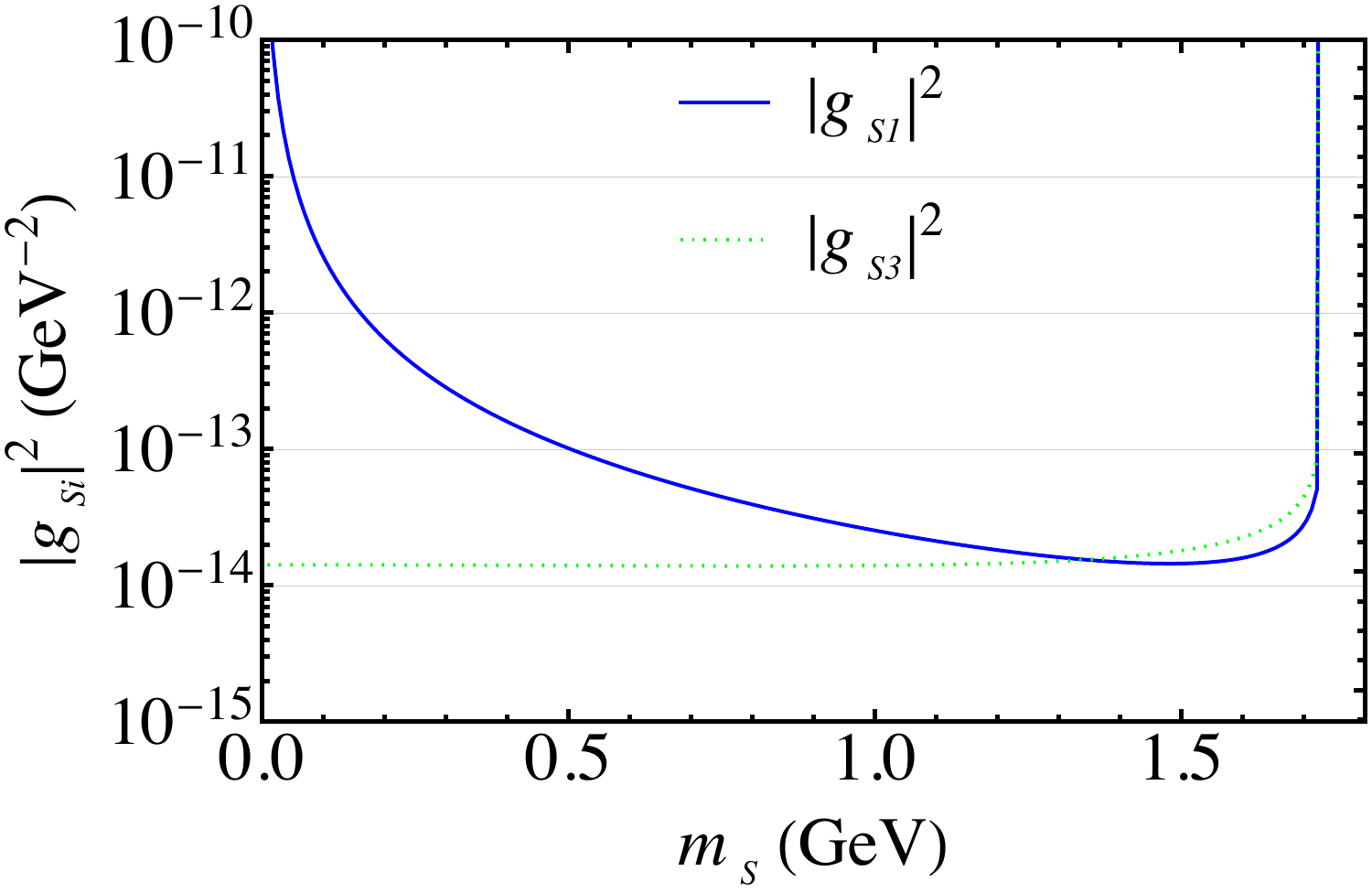}
	\caption{The upper limits on the coupling constants of $|g_{_{Si}}|^2$ as functions of $m_{_S}$.}
	\label{gs}
\end{figure}

\begin{figure}[htbp]
	\centering
	\subfigure[~$D^+\to\pi^+S$]{
		\label{br-D1pi1}
		\includegraphics[width=0.45\textwidth]{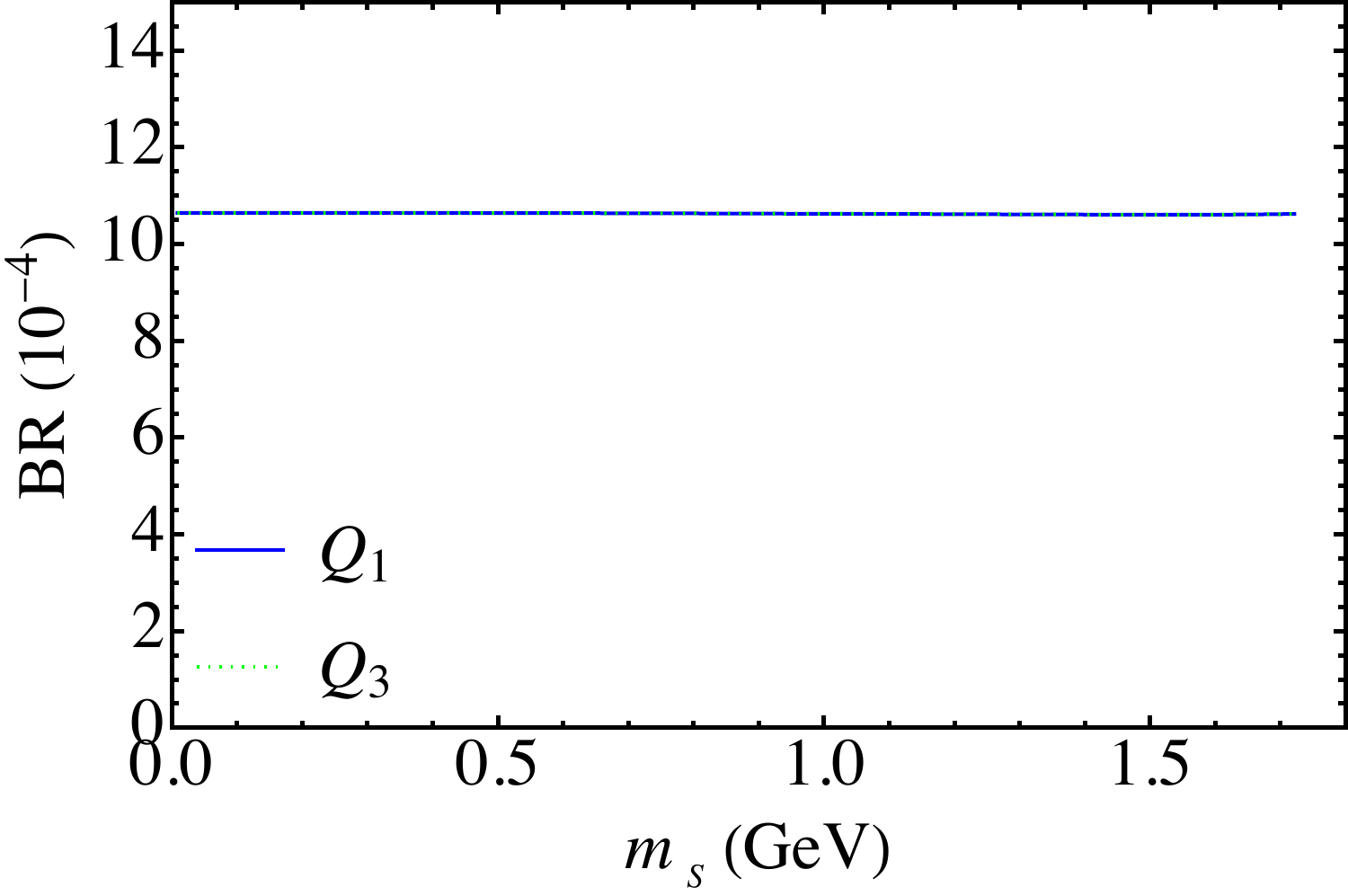}} 
	\hspace{2em}
	\subfigure[~$D_s^+\to K^+S$]{
		\label{br-DsK}
		\includegraphics[width=0.45\textwidth]{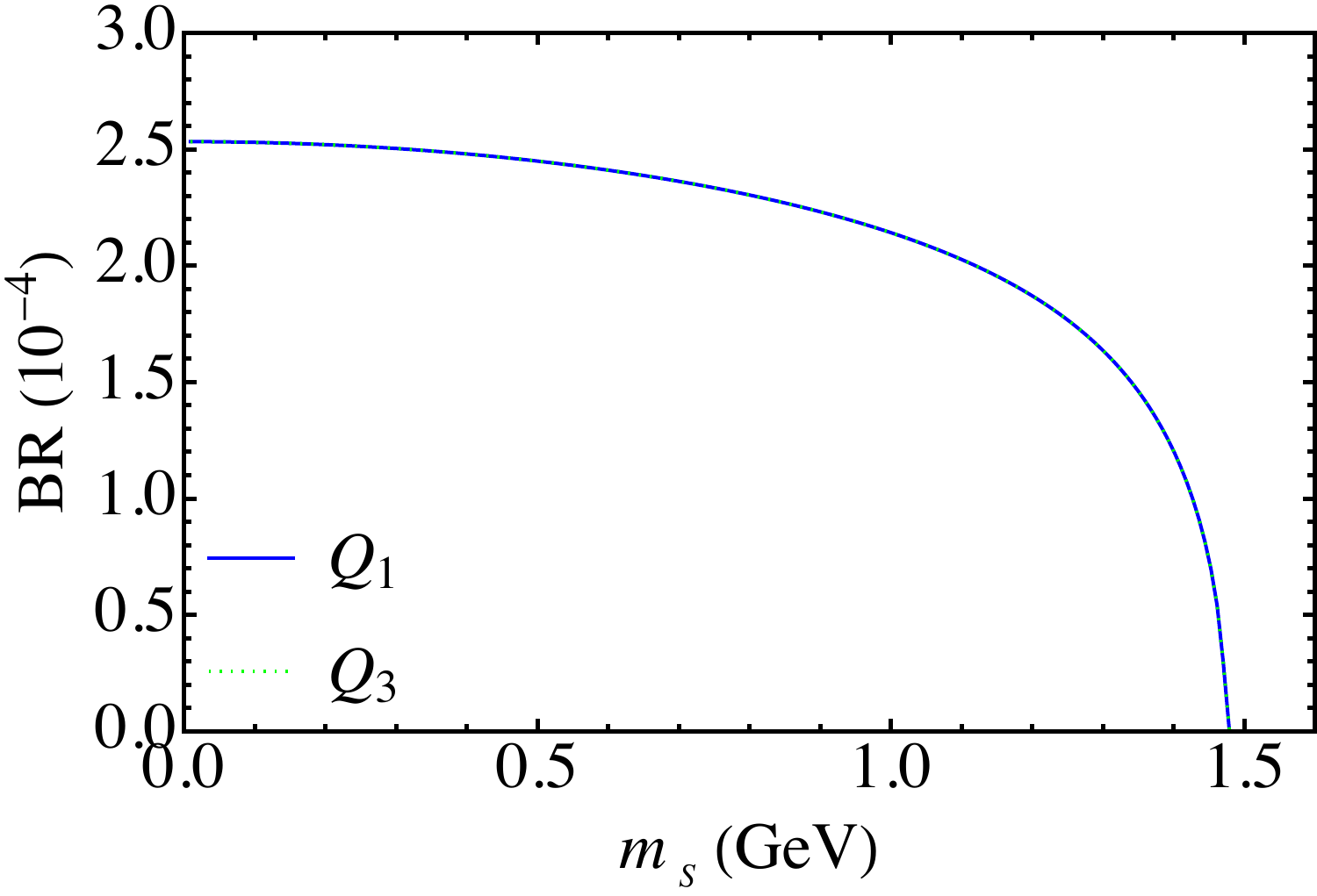}}\\
	\subfigure[~$\Lambda_c\to pS$]{
		\label{br-Lcp}
		\includegraphics[width=0.45\textwidth]{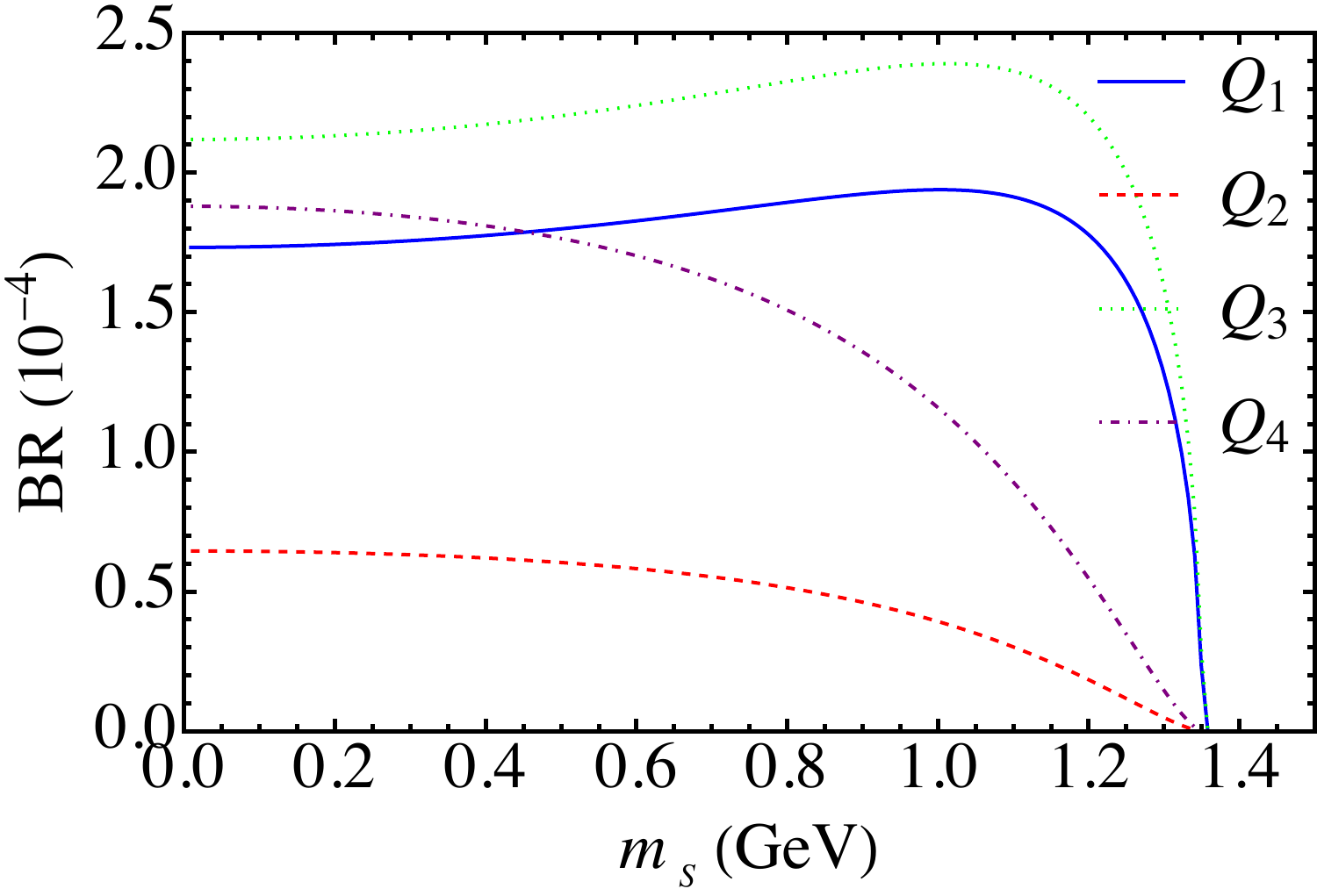}} 
	\hspace{2em}
	\subfigure[~$\Xi_c^{+}\to \Sigma^+S$]{
		\label{br-Xc1S1}
		\includegraphics[width=0.45\textwidth]{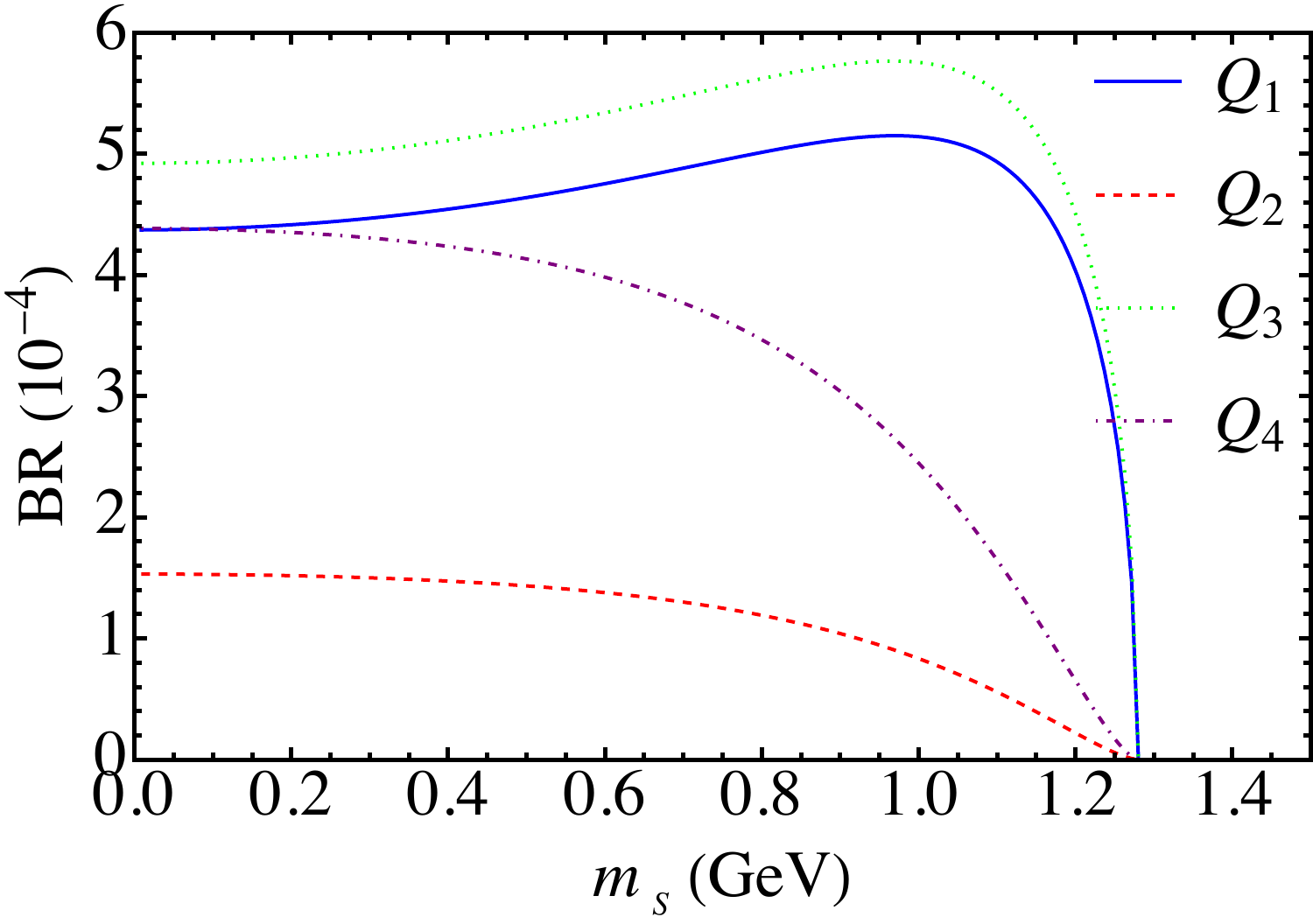}} \\
	\subfigure[~$\Xi_c^{0}\to \Sigma^0S$]{
		\label{br-Xc0S0}
		\includegraphics[width=0.45\textwidth]{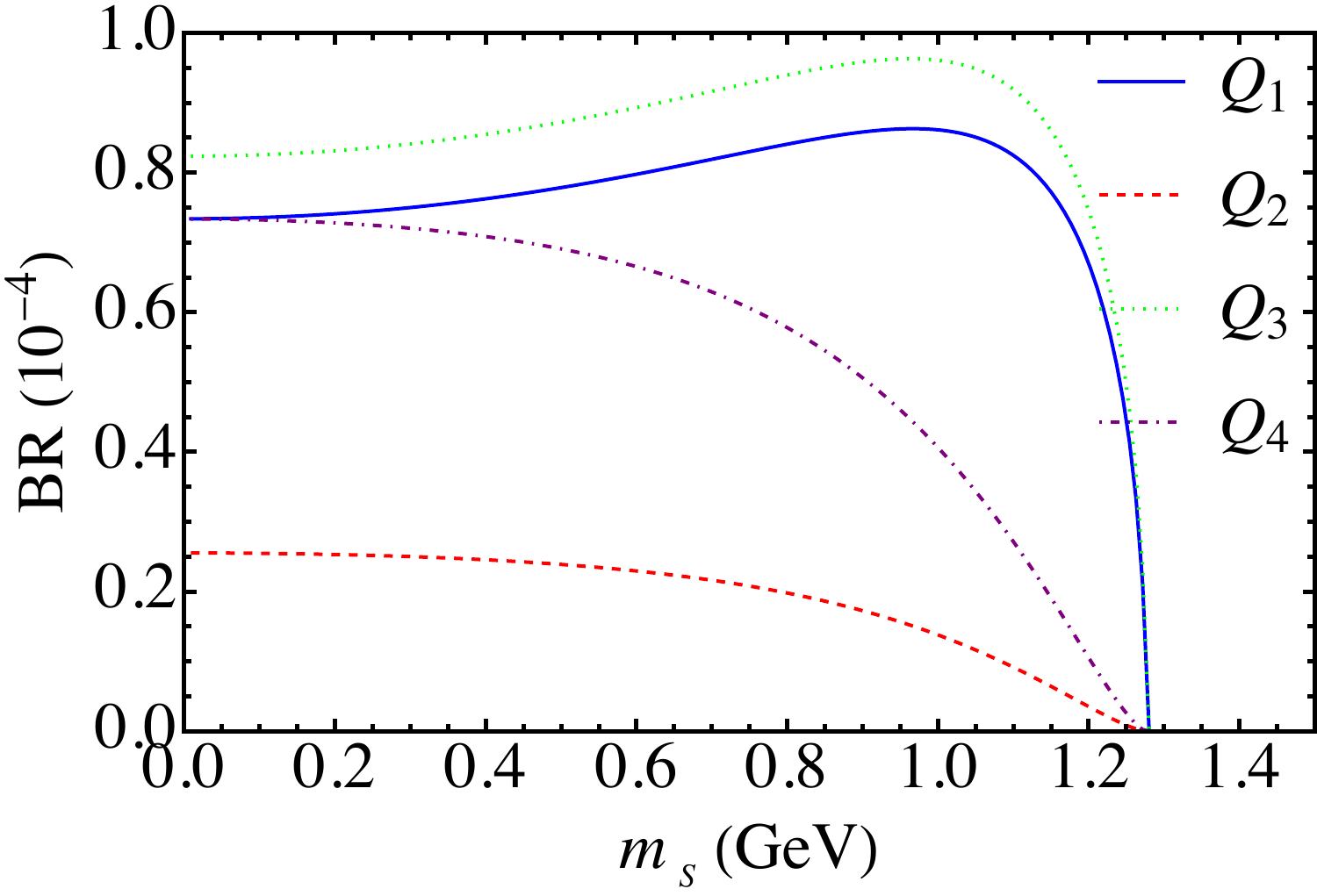}} 
	\hspace{2em}
	\subfigure[~$\Xi_c^{0}\to \Lambda S$]{
		\label{br-Xc0L}
		\includegraphics[width=0.45\textwidth]{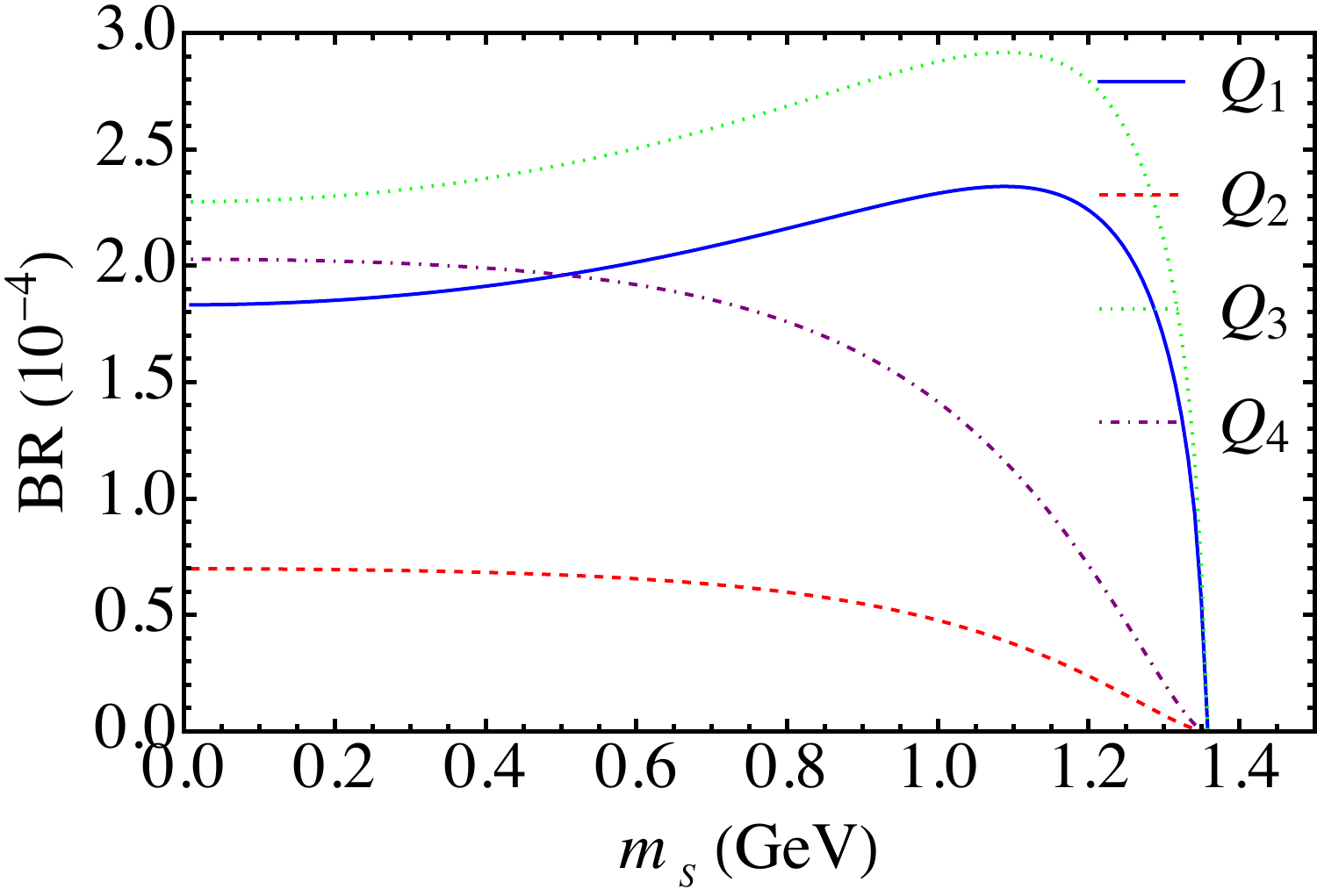}} 
	\caption{Upper limits of the branching ratios as functions of $m_{_S}$ in (a) $D^+\to\pi^+S$, (b) $D_s^+\to K^+S$, (c) $\Lambda_c\to pS$, (d) $\Xi_c^{+}\to \Sigma^+S$, (e) $\Xi_c^{0}\to \Sigma^0S$, and (f) $\Xi_c^{0}\to \Lambda S$.}
	\label{br}
\end{figure}

Currently, the only experimental upper limit on invisible FCNC processes of charmed hadrons is from the BES III experiment~\cite{BESIII:2021slf}, namely, $\mathcal {B}(D^0\to\pi^0 S) < 2.1\times 10^{-4}$. When we assume that only one operator contributes at a time, the upper bounds on coupling constants $|g_{_{11,33}}|^2$ as functions of $m_{_S}$ are obtained in Fig.~\ref{gs} by combining the experimental bound and our calculation. 
We find that in the most regions of $m_{_S}$, $|g_{_{11,33}}|^2$ are of the order of $10^{-14}$ to $10^{-13}$. When $m_{_S}\to0$, $|g_{_{S1}}|^2\to\infty$, because $\widetilde\Gamma_{11}$ is zero at this time. When $m_{_S}=(M-M_f)$, we find that the upper limits on $|g_{_{S1,S3}}|^2$ are infinite for the same reason. Since $\widetilde\Gamma_{11(33)}$ and $\widetilde\Gamma_{22(44)}$ have similar trends, the bounds on $g_{_{S1(3)}}$ can be extended to $g_{_{S2(4)}}$, respectively. With these bounds as inputs, we predict the upper limits on the branching ratios of $D^{+}\to\pi^{+}S$, $D_s^+\to K^+S$, $\Lambda_c\to pS$, $\Xi_c^{0(+)}\to \Sigma^{0(+)}S$, and $\Xi_c^0\to\Lambda S$, which are illustrated in Fig.~\ref{br}. 
Note that the contributions to the upper limits of the branching ratios by $Q_i$ are of the same order of $10^{-4}$. The bound on $\mathcal B(D^+\to \pi^+ S)$ is almost a horizontal line near $10.6\times 10^{-4}$, since we use the same form factors for $D^0\to K^0$ and $D^+\to K^+$ channels. The peak occurs around $m_{_{S}}=1.1 ~{\rm GeV}$, which is the region most likely to be detected experimentally. The branching ratios of $D^{+}\to\pi^{+}S$, $D_s^+\to K^+S$, $\Lambda_c\to pS$, $\Xi_c^{0(+)}\to \Sigma^{0(+)}S$, and $\Xi_c^0\to\Lambda S$ are mainly contributed by the operator $Q_3$, which can be as large as $10.6,~2.53,~2.39,~0.963,~5.77$, and $2.92\times10^{-4}$, respectively. When $m_{_S}=(M-M_f)$, they all become zero. These limits are much larger than the SM expectations of $\mathcal O(10^{-17})$ to $\mathcal O(10^{-16})$. We expect that the near-future experiments could give more results on the charmed hardron FCNC decays.

\section{Conclusion}
We have studied the light invisible scalar in the FCNC processes of the long-lived charmed mesons and baryons. The model-independent effective Lagrangian which contains four dimension-5 operators has been introduced to describe the couplings between the quarks and invisible scalar. The bounds of the coupling constants have been extracted from the recent BES III experiment on $D^{0}\to\pi^{0}\bar\nu\nu$. Based on these bounds, we have predicted the upper limits of $\mathcal B({\bf B}_c\to{\bf B}_u\chi\chi)$. In particular, we have found that the decay branching ratios of $D^{+}\to\pi^{+}S$, $D_s^+\to K^+S$, $\Lambda_c\to pS$, $\Xi_c^{0(+)}\to \Sigma^{0(+)}S$, and $\Xi_c^0\to\Lambda S$ can be as large as $10.6,~2.53,~2.39,~0.963,~5.77$, and $2.92\times10^{-4}$ with $m_{_S}\approx 1.1~{\rm GeV}$, respectively. We are looking forward to the future experimental searches, such as those at BES III, LHCb, Belle II and FCC-ee, to get more measurements on charmed hadrons to find signs of NP.

\section{Acknowledgments}
This work is supported in part by the National Key Research and Development Program of China under Grant No.~2020YFC2201501 and the National Natural Science Foundation of China (NSFC) under Grant No.~12147103 and No.~12247160.

\bibliography{reference}

\end{document}